\documentstyle[12pt]{article}
\catcode`\@=11
\@addtoreset{equation}{section}

\def\@normalsize{\@setsize\normalsize{15pt}\xiipt\@xiipt
\abovedisplayskip 14pt plus3pt minus3pt%
\belowdisplayskip \abovedisplayskip
\abovedisplayshortskip  \z@ plus3pt%
\belowdisplayshortskip  7pt plus3.5pt minus0pt}
\def\small{\@setsize\small{13.6pt}\xipt\@xipt
\abovedisplayskip 16pt plus3pt minus3pt%
\belowdisplayskip \abovedisplayskip
\abovedisplayshortskip  \z@ plus3pt%
\belowdisplayshortskip  7pt plus3.5pt minus0pt
\def\@listi{\parsep 4.5pt plus 2pt minus 1pt
            \itemsep \parsep
            \topsep 9pt plus 3pt minus 3pt}}
\def\underline#1{\relax\ifmmode\@@underline#1\else
	$\@@underline{\hbox{#1}}$\relax\fi}
\@twosidetrue
\catcode`@=12
\catcode`\@=11

\def\FERMIPUB{}
\def\FERMILABPub#1{\def\FERMIPUB{#1}}
\def\ps@headings{\def\@oddfoot{}\def\@evenfoot{}
\def\@oddhead{\hbox{}\hfill
	\makebox[.5\textwidth]{\raggedright\ignorespaces --\thepage{}--
	\hfill {\rm FERMILAB--Pub--\FERMIPUB}}}
\def\@evenhead{\@oddhead}
\def\subsectionmark##1{\markboth{##1}{}}
}
\ps@headings
\catcode`\@=12
\relax

\newskip\humongous \humongous=0pt plus 1000pt minus 1000pt

\newif\ifdtup

\def\oldreffmt#1{\rlap{[#1]} \hbox to 2\parindent{}}

\def\figfmt#1{\rlap{Figure {#1}} \hbox to 1in{}}

\def\etal{\hbox{\it et al.}}

\def\tr{\mathop{\rm tr}}
\def\Tr{\mathop{\rm Tr}}

\def\bra#1{\left\langle #1\right|}
\def\ket#1{\left| #1\right\rangle}

\def\ltap{\raisebox{-.4ex}{\rlap{$\sim$}} \raisebox{.4ex}{$<$}}

\def\beq{\begin{equation}}
\def\eeq{\end{equation}}
\def\bea{\begin{eqnarray}}
\def\eea{\end{eqnarray}}
\def\half{\frac{1}{2}}

\def\bq{\begin{quote}}
\def\eq{\end{quote}}

\def\half{\frac{1}{2}}
\def \lta {\mathrel{\vcenter
     {\hbox{$<$}\nointerlineskip\hbox{$\sim$}}}}
\def \gta {\mathrel{\vcenter
     {\hbox{$>$}\nointerlineskip\hbox{$\sim$}}}}
\def \etal {{\it et al.}\ }
\relax
\documentstyle[12pt]{article}
\begin{document}
\FERMILABPub{92/218}
\begin{titlepage}
\begin{flushright}
FERMILAB--Pub--92/218--T\\
October 1992\\
hep-ph/9210233\\
\end{flushright}
\vfill
\par \vskip .05in
\begin{center}
{\large \bf Spontaneously Broken Technicolor\\
and the
Dynamics of Virtual Vector Technimesons\\
}
 \end{center}
  \par \vskip .1in \noindent
\begin{center}
{\bf  Christopher T. Hill$^1$,
Dallas C. Kennedy$^1$\\ Tetsuya Onogi$^2$, Hoi--Lai Yu$^3$}
\end{center}
\begin{center}
  \par \vskip .1in \noindent
{$^1$ Fermi National Accelerator Laboratory\\
P.O. Box 500, Batavia, Illinois, 60510, USA}
  \par \vskip .1in \noindent
{$^2$ Department of Physics\\
Hiroshima University, 1-3-1 Kagamiyama\\
Higashi--Hiroshima, 724, { Japan}}
  \par \vskip .1in \noindent
{$^3$ Institute of Physics\\
Academia Sinica\\ Nankang, Taipei, 11529, Taiwan, R.O.C.}
  \par \vskip .1in \noindent
\end{center}
\begin{center}{\large Abstract}\end{center}
\par \vskip .1in
\begin{quote}
We propose spontaneously breaking technicolor, thus
liberating techniquarks and suppressing large resonance contributions
to the electroweak $S$ parameter.
The dynamics is modeled by a fermion bubble
approximation to a single massive technigluon exchange
potential.
This contains a Nambu--Jona-Lasinio
model with additional interactions.  ``Virtual'' vector mesons occur
and contribute to $S$, and their
effects are studied.
Models of broken technicolor are discussed.
\end{quote}
\vfill
\end{titlepage}

\noindent
{\bf I. Introduction}
\vskip .1in

Technicolor is the collective name for a class of models attempting to
explain electroweak symmetry breaking by condensates of
fermion-antifermion pairs, driven by a new
strong interaction [1].
All versions  of technicolor (TC) imitate, more or less, the known
dynamics of QCD chiral symmetry
breaking. The technicolor interaction
is assumed to be an unbroken, and thus confining,
strong interaction amongst techniquarks, which also carry
electroweak quantum numbers.  The resulting
chiral condensate of techniquarks then breaks the electroweak
gauge symmetry.  The extension
of the theory to give the fermions masses [2] is made
somewhat awkward by the twin constraints of flavor-changing
neutral current processes and the large
top quark mass.
Implementing these constraints in extended technicolor (ETC) leads to
additional model building requirements, such
as a walking technicolor coupling [3] and critical or subcritical
extended technicolor [4].  The mass of the top quark
can then be generated either by a ``subcritical amplification''
involving tuning of the ETC four--fermion interactions near to, but below, the
critical value, or by generating techniquark and top quark
condensates together, through ETC effects.

There is a second difficulty which appears to imply
that technicolor theories are no longer viable.
Techni-resonances such as the
techni--$\rho$ and techni--$A_1$, arising from confinement
of techniquarks,
produce large contributions to the electroweak radiative $S$ parameter,
a measure of isosinglet electroweak symmetry breaking at the loop level [5,6].
In a QCD--like theory these can be treated by
resonance saturation of superconvergence relationships, such
as the Weinberg sum rules [6].  Recently, precise measurements of electroweak
interactions have placed a new constraint
on the  $S$ parameter, currently at 90\% (95\%) to $< -0.1\ (0.0)$ [7],
relative
to the minimal Standard Model with the Higgs mass equal to the $Z$ boson mass.
In technicolor theories treated as scaled-up QCD--like theories,
we have:
\beq
S \simeq (0.10)N_{TC}N_{TD} + 0.13,
\eeq
where the two indices are the number of technicolors and electroweak
technidoublets, respectively [6].  In walking technicolor theories,
\beq
S \simeq (0.11a - 0.07b)N_{TC}N_{TD} + 0.13,
\eeq
where $a,b$ are constants of order unity [8].  The final contribution
in both cases comes from taking the Higgs mass to $\sim$ 1 TeV.
By comparison, the perturbative form of $S$ from fermion loops
alone [5] is:
\beq
S = N_{TC}N_{TD}/6\pi \approx (0.05)N_{TC}N_{TD}.
\eeq

If technicolor theories are to be
viable, their contribution to $S$ must be minimized
somehow. In the present article, we wish to present an alternative
realization of technicolor as a spontaneously broken gauge theory and
study to what
extent this can suppress techni--resonance  contributions to the $S$ parameter.
In such a theory, the effective
coupling constant must be just large enough to drive the formation of
electroweak condensates, but the technicolor interaction kept
short-range, approximately $s$--wave and not confining.
While scalar bound states such as the Nambu--Goldstone
bosons are formed, the unconfined
technifermions
do not form the offending vector resonances which give large  contributions
to the coefficients appearing in $S$.  There are, however, ``virtual''
resonance poles formed at a scale above the breaking scale.  The effects
of these virtual resonances upon $S$ must be included. We
see below, in a large--$N$ analysis,
how the intuition of a suppressed $S$ in
spontaneously broken technicolor is achieved.
While we find that there is suppression, $S$ never falls below the usual result
for free fermion loops, within the domain of validity of our analysis.

The  proposal of a spontaneously broken
technicolor (SBTC) is not really so radical in light of other recent
ideas about electroweak symmetry breaking.
Another alternative for dynamical electroweak
symmetry breaking is to abandon new technifermions
and suppose that the top quark itself forms a
condensate and acts alone as a techniquark [9].  A preliminary
gauge form of this dynamics has been given [10], and the requirement that the
electroweak $\rho$ parameter be approximately unity
forces some radical
fine-tuning: either (a) the scale of new physics $\Lambda$ is
very large, e.g., $10^{15}$ GeV and the
fine-tuning very severe for the effective quadratic
interactions, or (b) supersymmetry must be invoked,
or (c) $m_{top}>>200 $ GeV
and an {\it ad hoc} cancellation must be
arranged against the large positive top quark contribution to $\rho$, such as
a new heavy $Z'$ or a new broken $SU(2)_V$.  These problems disappear
if a fourth generation is invoked [11].
In these cases, the number of degrees of freedom
contributing to $S$ is minimized.
Moreover, in these models we are dealing with
a new strong
interaction which is itself broken and does not confine, but which is
sufficiently strong to be near or beyond criticality.

We cannot solve strong coupling models without resort to some
approximation, such as the leading--$N$ fermion bubbles,
or ladder approximations.
Our view of the dynamics is that a given
SBTC gauge theory
is described by a gauge group, $G_{SBTC}$. The theory must be
asymptotically free at high energies.  If we turn off the breaking mechanism,
then it becomes a confining theory with an infrared confinement scale
$\Lambda_{SBTC}$.  However, we imagine that some mechanism intervenes to
break $G_{SBTC}\rightarrow G'$ at a scale $\Lambda^\prime
 \geq \Lambda_{SBTC}$.  We assume
to start off that $G'$
is a null group (that is, complete breaking; we could generalize to have
unbroken subgroups that contain $U(1)$ factors or are infrared--free).
Thus with $G'$ null,
all of the technigluons acquire a mass $M \sim g(M)\Lambda^\prime ,$
where $g(\mu )$ is the $G_{SBTC}$ coupling and $g(M)\sim {\cal O}(1-10).$

At the scale $M$ we can integrate out the gauge bosons and replace the
gauge interactions with four-fermion effective interactions.  Thus at and
below this scale we have a Nambu--Jona-Lasinio (NJL) model [12].  Our criterion
for the symmetry breaking of the theory we take for the sake of simplicity
to be the usual fermion bubble, Nambu--Jona-Lasinio result, that
a chiral symmetry breaking scale or mass gap $m$ be
spontaneously generated for sufficiently strong
coupling.    We then study the
effects of the virtual resonances $\rho$ and $A_1$ in the model
in leading order of $1/N_{TC}$.
This may be viewed as a model calculation which seeks
to understand directly the virtual resonance contributions to $S$
without use of the Weinberg sum rules.  We find that
the resonance contribution to $S$ is always
positive, and in the limit of tuning a large hierarchy between $M$ and the
chiral beaking scale $m$, we recover the usual free fermion result.

\vskip .1in
\noindent
{\bf II. A Simple Model of the Dynamics}
\vskip .1in

Consider an SBTC theory, with a single fermionic flavor isodoublet ($N_{TD}$ =
1,
$N_{TF}$ = 2) of the form $\psi = (U,D)$ which also carries an $SU(N_{TC})$
color index in the fundamental representation. We do not presently
concern ourselves with anomalies.
We  assume that the  $SU(N_{TC})$ local gauge theory is
broken to a global $SU(N_{TC})$ with $N_{TC}$ colors.
(Our discussion can be generalized to $N_{TD} > 1$, and other breaking
schemes are mentioned in section IV.)
We can write the effective current--current form of the fermion
interaction Lagrangian due to single gluon exchange of momentum
transfer $q^2 << M^2$ as:
\beq
{\cal L}_{\rm int} = -\frac{g^2}{M^2}\bar{\psi}
\gamma_\mu\frac{\lambda^A}{2}\psi\bar{\psi}
\gamma^\mu\frac{\lambda^A}{2}\psi ,
\eeq
where the $\lambda^A$ are the
$SU(N_{TC})$ generators.
Upon Fierz rearrangment, keeping only leading terms in
$1/N_{TC}$, this interaction takes the form:
\begin{eqnarray}
{\cal L}_{\rm int}   & = &
\frac{g^2}{M^2}\left(
\bar{\psi}_L\psi_{R}\bar{\psi}_{R}\psi_{L} +
\bar{\psi}_L\tau^a\psi_{R}\bar{\psi}_{R}\tau^a\psi_{L}\right.
\nonumber \\
& &
-\frac{1}{8}\bar{\psi}\gamma_\mu\tau^a\psi \bar{\psi}\gamma^\mu\tau^a\psi
-\frac{1}{8}\bar{\psi}\gamma_\mu\gamma_5\tau^a\psi
\bar{\psi}\gamma^\mu\gamma_5\tau^a\psi
\nonumber \\
& & \left.
-\frac{1}{8}\bar{\psi}\gamma_\mu\psi \bar{\psi}\gamma^\mu\psi
-\frac{1}{8}\bar{\psi}\gamma_\mu\gamma_5\psi \bar{\psi}\gamma^\mu\gamma_5\psi
\right),
\end{eqnarray}
where $\psi_L = (1-\gamma_5)\psi/2$,
$\psi_R = (1+\gamma_5)\psi/2$.
Here $\tau^a$ are Pauli matrices acting upon the flavor isospin indices.
Notice the $SU(2)_L\times SU(2)_R$ invariance of this interaction, where
the full chiral group has an $SU(2)$ vector custodial subgroup.
The first two terms are Nambu--Jona-Lasinio interactions, and $M^2$ plays the
role of the cutoff.  (This can be rigorously checked by comparing
arguments of logs and finite corrections in various amplitudes;
it is generally found that $M^2$ can be identified with the NJL cutoff
with small residual corrections.) The vector-vector and
axial-vector--axial-vector terms generate the resonance contributions
to the $S$ parameter.

We begin by demanding that the theory produce a vacuum condensate at
low--energy,
or equivalently, a dynamically induced effective fermion mass.  The
most general induced mass term can be taken to be:
\beq
 m \bar{\psi}\psi + \delta m \bar{\psi}\tau^3\psi .
\eeq
The fermion mass is generated self-consistently by its own self-energy in the
NJL sector of the theory.  This consists of the first two
terms of the {\em rhs} of eq.(5).   In the leading--$N$
approximation, only the NJL interactions can
contribute to the mass gap.
The resulting gap equations for $m$ and $\delta m$ are
most conveniently written in terms of $m_\pm = m \pm \delta m$:
\beq
m_\pm = \frac{g^2(M^2)N_{TC}}{8M^2\pi^2}\left[
m_+(M^2 - m_+^2\ln(M^2/m^2_+ + 1)) +\\
m_-(M^2 - m_-^2\ln(M^2/m^2_- + 1))
\right] .
\eeq
We therefore see that
$\delta m = 0$, and the symmetric mass $m$ satisfies the NJL gap equation:
\beq
m = \frac{g^2(M^2)N_{TC}}{4M^2\pi^2}m\left[
M^2 - m^2\ln(M^2/m^2 + 1)
\right] .
\eeq
We ignore the trivial solution $m = 0.$
Thus, the condition that a chiral condensate form in the theory
at scale $M$ is the usual condition
in the Nambu--Jona-Lasinio model:
\beq
g^2(M^2) = 4\pi^2\eta/N_{TC} \geq 4\pi^2/N_{TC} ,
\eeq
or $\eta\geq 1,$
insofar as the large--$N$ limit is valid.  For marginally critical coupling,
$\eta
\rightarrow 1^+$ and $m\rightarrow 0;$ while for large $\eta ,$
$m\rightarrow\infty .$  We implicitly assume $m\leq M$ for an effective field
theory, so that $\eta\ltap$ 3.3.  We also note that the argument of
the logarithms $(1+ M^2/m^2)$, can be replaced by  $M^2/m^2$, since upon
expandi
ng,
the corrections are higher order in $m^2/M^2$, and we have already truncated
operator corrections of this order in writing eq.(4).  This implicitly
requires $m$ to be small compared to $M$ and $\eta$ not much larger
than unity.
Note that
the result $\delta m$ = 0 is an example of the Vafa--Witten theorem [13],
that vector symmetries cannot be dynamically broken.  In this case, $\delta m$
= 0 preserves the vector custodial SU(2).

Let us argue now on general theoretical grounds that there
can exist a spontaneously broken, asymptotically free, theory that is
unconfining yet sufficiently strong to form a chiral condensate. If we take
$g^2(\mu)$ to be the running coupling constant on scales $\mu >>M$,
\beq
g^2(\mu) = \frac{8\pi^2}{b_0\ln(\mu/\Lambda_{SBTC})},
\eeq
then, if we assume $g^2(M)$ is marginally critical,
we have the ratio of $M$ to $\Lambda_{SBTC}$ given by:
\beq
\frac{M}{\Lambda_{SBTC}} = \exp(2N_{TC}/b_0)
\eeq
where
\beq
b_0 = \frac{11N_{TC}}{3}- \frac{2N_{TF}}{3}.
\eeq
Therefore, we can, by judicious choice of $N_{TC}$ and $N_{TF}$,
make the ratio $M/\Lambda_{SBTC}$ arbitrarily large.  This includes
a ``walking theory'' in which $b_0\sim 0$, or
$N_{TF} \sim 11N_{TC}/2$.  For $N_{TC}$
very large, we then expect the Nambu--Jona-Lasinio
approximation to the chiral dynamics to be very good.
Of course, this merely demonstrates that such an SBTC theory can exist
as a matter of principle, while our specific model results are expected to be
less reliable.

Since the  SBTC condensate breaks the
electroweak gauge symmetry, there are two charged and one neutral
Nambu--Goldstone bosons [12,14] which become
the longitudinal $W$ and $Z$ bosons respectively.
The decay constants of these Goldstone bosons
are denoted  by $f_W(p^2)$ and $f_Z(p^2)$
respectively, being generally functions
of momentum transfer $p^2$.
The decay constants occur in writing the vacuum polarization
tensors for two-point functions of electroweak currents.
Let us define:
\beq
\Pi_{\mu\nu}(p^2) = (g_{\mu\nu} - p_\mu p_\nu/p^2)\Pi(p^2).
\eeq
Then we have (in the convention of writing kinetic terms
for gauge fields as $(-1/4g^2)F_{\mu\nu}F^{\mu\nu}$, e.g. see ref.[9]):
\bea
\Pi_{\pm}(p^2) & = & \left( \frac{1}{g_2^2}p^2 - f_W^2 \right)
\nonumber \\
\Pi_{33}(p^2) & = & \left( \frac{1}{g_2^2}p^2 - f_Z^2 \right)
\nonumber \\
\Pi_{3Q}(p^2) & = & \left( \frac{1}{g_2^2}p^2 \right)
\eea
where $g_2$ is the $SU(2)_L$ gauge coupling.
These are left--handed current--current two point functions,
with $\half(1-\gamma^5)$ projections.
Since custodial $SU(2)$ is unbroken, we have
$m_+ = m_-$ and  $\Pi_{\pm} = \Pi_{33}$, whence:
\bea
f_W^2 =  f_Z^2 & = &  \Pi_{33} - \Pi_{3Q}
\\
& = &  -\frac{1}{4}[\Pi^{VV}_{33} - \Pi^{AA}_{33}],
\eea
where use has been made of $Q = I_{L3} + \frac{Y}{2}$,
and $VV$ ($AA$) are vector (axial vector) two--point functions,
and the factor of $1/4$ arises from the $\half$ in the left--handed
projections, $\half(1-\gamma^5)$.

Let us make some preliminary comments about the physical meaning
of the $S$, $T$ and $U$ parameters.
In general, we can expand $f_W^2(p^2)$ and $f_Z^2(p^2)$ in a Taylor
series in $p^2$:
\begin{eqnarray}
f_W^2(p^2) &=& f_0^2 +  \half\sigma p^2 +  \half\tau f_0^2  +  \half\omega p^2
+ \;...
\\
f_Z^2(p^2) &=& f_0^2 +   \half\sigma p^2 -  \half\tau f_0^2  -  \half\omega p^2
+\;...\;.
\end{eqnarray}
$f_0$  is just the Higgs vacuum expectation
value in the standard model.
We use a normalization in which $f_W^2(0) = 1/4\sqrt{2}G_F$,
or $f_W(0)\approx 123$ GeV.
Furthermore, note that the (isospin-breaking) $\tau$
parameter is just a rewriting
of the $\rho$ parameter, since $\rho = f_W^2(0)/f_Z^2(0)$. The
parameters $\sigma$ and $\omega$ are the isospin-conserving and
isospin-breaking measures respectively
of physics contributing to the $p^2$ evolution of
the low-energy effective theory.  The $p^2$ expansion
about zero is
strictly valid only for heavy
contributions to the $f_X^2(p^2),$ because singularities would occur
for, e.g., massless neutrinos which give
$\sim \ln p^2 $ terms in $\sigma$,
$\tau$ and $\omega$.
However, the physical electroweak
observables actually depend on the $f_X(p^2),$ not on their derivatives, and
the $f_X(p^2)$ are not singular,
only their derivatives are at $p^2 \rightarrow 0$.

With the conventional definitions of $S$, $T$ and $U$ [6]:
\begin{eqnarray}
S &= & 16\pi\frac{\partial}{\partial p^2}\left[\Pi_{33}
 -          \Pi_{3Q}\right]_{p^2=0}
\\
T &= & \frac{4\pi}{\sin^2\theta\cos^2\theta M_Z^2}\left[ \Pi_{\pm}
 -          \Pi_{33}\right]_{p^2=0}
\\
U &= & 16\pi\frac{\partial}{\partial p^2}\left[\Pi_{\pm}
 -          \Pi_{33}\right]_{p^2=0},
\end{eqnarray}
we can obtain the following relations to the decay constant
parameters introduced above:
\beq
\sigma = \frac{2S+U}{16\pi},\qquad
\omega = \frac{U}{16\pi},\qquad
\tau = \frac{\sin^2\theta\cos^2\theta M_Z^2T}{4\pi v_r^2}
= \frac{\alpha}{2}T.\qquad
\eeq
Note that in the limit of exact custodial $SU(2)$ symmetry, $T=U=0$, and
$S$ is equivalent to $\sigma$ and parameterizes the $p^2$ evolution
of $f^2$ in the theory.

Let us now focus upon the quantity:
\begin{eqnarray}
f^2(p^2)  & = &
 -\frac{1}{4}\left[\Pi^{VV}_{33} -  \Pi^{AA}_{33}\right] ,
\end{eqnarray}
defined in terms of the vector and axial-vector current correlators.
First we compute $\Pi^{VV}_{33}$:
\begin{eqnarray}
[\Pi_{33}^{VV}](p^2)_{\mu\nu} & = &
i\bra{0}T\; \bar{\psi}\gamma_\mu\frac{\tau^3}{2}\psi \;
\bar{\psi}\gamma_\nu\frac{\tau^3}{2}\psi \ket{0}
\nonumber \\
& = &
\frac{N_{TC}}{4\pi^{2}}\int_0^1\; dx
\left[x(1-x)(g_{\mu\nu}p^2-p_\mu p_\nu)\right]
\nonumber \\ & &
\qquad \times \ln\left\{M^2/ (m^2-x(1-x)p^2 )\right\}.
\end{eqnarray}
Note that this expression is transverse, as it should be,
owing to the conserved vector current (CVC),
$p^\mu\bar\psi\gamma_\mu(\tau_3/2)\psi$ = 0.
Now we compute $\Pi_{AA}^{\mu\nu}$:
\begin{eqnarray}
[\Pi_{33}^{AA}](p^2)_{\mu\nu} & = & i\bra{0}T\; \bar{\psi}\gamma_\mu
\gamma_5\frac{\tau^3}{2}\psi \; \bar{\psi}\gamma_\nu
\gamma_5\frac{\tau^3}{2}\psi
\ket{0}
\nonumber \\
& = &
 \frac{N_{TC}}{4\pi^{2}}\int_0^1\; dx
\left[x(1-x)(g_{\mu\nu}p^2-p_\mu p_\nu) - g_{\mu\nu} m^2 \right]
\nonumber \\ & &
\qquad \times \ln\left\{M^2/ (m^2-x(1-x)p^2 )\right\}.
\end{eqnarray}
This expression is not transverse, owing to the dynamical symmetry
breaking ($m\neq 0$).
However, if we now sum the leading
large--$N_{TC}$ effects of the NJL interactions,
and make use of the gap equation, we form the Goldstone pole
in the usual way, and
then the full amplitude becomes transverse:
\begin{eqnarray}
[\widetilde\Pi^{AA}_{33}](p^2)_{\mu\nu} & = & \frac{N_{TC}}{4\pi^{2}}
(g_{\mu\nu}-\frac{p_\mu p_\nu}{p^2})\int_0^1\; dx
\left[x(1-x)p^2 -  m^2 \right]
\nonumber \\ & &
\qquad \times \ln\left\{M^2/ (m^2-x(1-x)p^2 )\right\} .
\end{eqnarray}

These expressions contain the usual
pure Nambu--Jona-Lasinio result for the
decay constant:
\begin{eqnarray}
f_{}^2(p^2) & = & -\frac{1}{4}\left[\Pi^{VV}_{33} -
\widetilde{\Pi}^{AA}_{33}\right]
\nonumber \\
 & = &  \frac{N_{TC}}{16\pi^{2}}\int_0^1\; dx
(m^2)
\ln [M^2 / (m^2 - x(1-x)p^2) ] .
\end{eqnarray}
It is instructive to write the expressions for general $m_\pm $ [9]:
\begin{eqnarray}
f_{Z0}^2(p^2) & = & \half N_{TC}m_{+}^2(4\pi)^{-2}\int_0^1\; dx
\ln [ M^2/(m_{+}^2 - x(1-x)p^2) ]
\nonumber \\
& & +\half N_{TC}m_-^2(4\pi)^{-2}\int_0^1\; dx
\ln [M^2/(m_-^2 - x(1-x)p^2)]
\nonumber \\
f_{W0}^2(p^2) & = & N_{TC}(4\pi)^{-2}\int_0^1\; dx
(x m_+^2 + (1-x) m_-^2)
\nonumber \\
& & \qquad \times
\ln [M^2 /( xm_+^2 + (1-x)m_-^2- x(1-x)p^2) ] .
\end{eqnarray}
We write $f_{W0}^2$ and $f_{Z0}^2$ to denote that these quantities
are obtained in the NJL approximation.
Expanding in $p^2$
and extracting the coefficients we find
for the $S$ parameter:
\begin{eqnarray}
S_0 & =  & \frac{N_{TC}}{6\pi}
\left[1 - \frac{1}{3}
\ln\left\{ m_+^2 / m_-^2\right\}\right] ,
\end{eqnarray}
while $T$ is, modulo an overall factor,
just the usual Veltman expression for $\delta \rho$:
\begin{eqnarray}
T_0 &=&  \frac{N_{TC}}{4\pi \sin^2\theta \cos^2\theta M_Z^2}
\left[m_+^2+m_-^2
-\frac{2m_+^2m_-^2}{(m_+^2-m_-^2)}\ln(m_+^2/m_-^2)
\right] .
\end{eqnarray}
This is the standard result for free fermion loops.
Hence,  a spontaneously broken technicolor produces
in the NJL approximation the usual free fermion loop
result for $S$.  We note that for $m_+ = m_-,$ $T = U =0$.

Next, we include the additional vector-vector and axial-vector--axial-vector
terms of the full interaction Lagrangian.
These are generated when one performs bubble sums of
the
vector-vector and axial-vector--axial-vector  terms.
The full bubble sum
of the vector-vector interaction yields:
\begin{eqnarray}
[\overline\Pi^{VV}_{33}](p^2)_{\mu\nu} & = & \frac{N_{TC}}{4\pi^{2}}
(g_{\mu\nu}-\frac{p_\mu p_\nu}{p^2})
\\
& & \times
\frac{\int_0^1\; dx \;
x(1-x)p^2 \ln\left\{M^2/ (m^2-x(1-x)p^2 )\right\}}
{\left[1 - (G N_{TC}/4\pi^2) \int_0^1\; dx \;
x(1-x)p^2 \ln\left\{M^2/ (m^2-x(1-x)p^2 )\right\}\right]} ,
\nonumber
\end{eqnarray}
where $G = g^2(M^2)/M^2.$

We also sum the full axial-vector amplitudes
(note that this is a double summation, of those interactions
producing the Goldstone pole together with the axial-vector--axial-vector
vertices):
\bea
[\overline\Pi^{AA}_{33}](p^2)_{\mu\nu} & = & \frac{N_{TC}}{4\pi^{2}}
(g_{\mu\nu}-\frac{p_\mu p_\nu}{p^2})
\\
& &
\!\!\!\!\!\!\!\!\!\!\!\!\!\!\!\!\!\!\!\!\!\!\!\!\times
\frac{\int_0^1\; dx \;
(x(1-x)p^2 - m^2)\ln\left\{M^2/ (m^2-x(1-x)p^2 )\right\}}
{\left[1 - (G N_{TC}/4\pi^2)\int_0^1\; dx \;
(x(1-x)p^2 - m^2)\ln\left\{M^2/ (m^2-x(1-x)p^2 )\right\}\right] } .
\nonumber \eea

We  now observe that the theory
generates {\em virtual} vector meson poles.
We refer to these as {\em virtual} resonances
because the poles occur at $p^2 > M^2$, which is beyond the domain
of validity of the effective theory.  For $p^2 > M^2,$ there are only
quasi-free technifermions and no bound states.
Nonetheless, the effects
of these analogue resonances are real on scales $p^2 < M^2,$
and they give a nontrivial result for $S$.

We can consider the ratio of the denominators of eqs.~(31) and (32)
at $p^2=0$ as a definition of the ratio of the off-shell mass-squares of the
vector ($\rho$) and axial-vector ($A_1$) resonances, $m_\rho^2/m_{A_1}^2$:
\beq
\frac{m_\rho^2}{m_{A_1}^2} = \frac{M_0^2}{M_0^2 + m^2\ln(M^2/m^2)}
\eeq
where  $M_0^2 = 4\pi^2/GN_{TC}$.
We note that:
\beq
\eta = \frac{N_{TC} G M^2}{4\pi^2}  = M^2/M_0^2.
\eeq
Recall that $\eta=1$ corresponds to the critical coupling and that condensation
requires $\eta\ge$ 1.  The gap equation then states:
\beq
 1 = \eta \left[ 1 - \frac{m^2}{M^2}\ln(M^2/m^2)\right]
\eeq
so we obtain:
\beq
\frac{m_\rho^2}{m_{A_1}^2} = \frac{1}{\eta} .
\eeq
Our model does not give the usual Weinberg sum rule result for QCD,
that this ratio is one-half,
nor should it.  We do not have real $\rho$ and $A_1$ resonances,
and we cannot argue that $\Pi_{VV}-\Pi_{AA}$ is saturated by resonances.
We should not expect the present model, which captures the features of
chiral symmetry breaking  dynamics, to be applicable in general
to QCD.  In particular, the local interaction does not confine.  Thus the
model cannot be taken as very accurate for the higher (e.g., vector) states
of QCD.  On the other hand, the four-fermion interaction
eqs.(4,5) is {\em exact} in
the broken case, subject only to the low-energy and large--$N$ approximations.
So we should expect the results to be reasonable for broken technicolor.

Now we compute $S$:
\bea
S & = & -4\pi\frac{\partial}{\partial p^2}
\left[ \overline\Pi_{VV} - \overline\Pi_{AA} \right]_{p^2=0}
\nonumber \\
& = & \frac{N_{TC}}{6\pi}
\left[1 + (\ln(M^2/m^2) - 1)(1 - \frac{1}{\eta^2})
  \right]
\eea
where we note that the $\ln (M^2/m^2)$ is a function of
$\eta$ by the gap equation~(8, 35).  This relation is plotted in
Figure~1, and we discuss it in the Conclusions.
Notice that when we tune the theory near to criticality,
$\eta\rightarrow 1^+,$ and $S$ reduces
to the conventional free fermion loop result, $N_{TC}/6\pi$.  The model
calculation becomes unreliable as the logarithm becomes small, because the
mass gap $m$ approaches the cutoff $M.$ Notice that with $m<M$ the
parameter $S$ does not drop below the free fermion loop value.

\vskip .1in
\noindent
{\bf III. Effective Lagrangian Approach}
\vskip .1in

It is instructive to analyze the virtual vector meson effects by
way of an effective action approach.
We rewrite the four-fermion interaction Lagrangian by
introducing auxiliary fields which correspond to Higgs, techni-$\rho$ and
techni-$A_{1}$ states. These auxiliary fields are
non-dynamical at the scale $M$, having no kinetic terms,
but they  become dynamical fields from the effect of
the fermion loops as we evolve the effective action
to scales $\mu < M$.
We cannot  argue that these are
real propagating fields since the energy scale of the poles for these
virtual resonance
states is as large as the cutoff scale $M$. But
in computing $S$ one is
interested in the low--energy effect of the virtual resonance
kinetic terms, and one can reliably obtain
the correct answer in this approach.
Hence,  in this section we briefly explain how to obtain $S$ in the effective
action method. We  proceed in two steps.
First, we introduce auxiliary fields to rewrite the Lagrangian.
By integrating out the techni-fermion degrees of freedom, we obtain the
effective action for the Higgs, techni-$\rho$, techni-$A_{1}$, and
$SU(2)_L \times U(1)_Y$ gauge fields $W_{\mu}$ and $B_{\mu}$.
In the second step, we eliminate the techni-$\rho$ and techni-$A_{1}$
using the equation of motion. In this way we obtain the low--energy effective
Lagrangian for the $SU(2)_L \times U(1)_Y$ gauge fields. The $S$ parameter is
easily
read off the $W_{\mu}^{3}$ - $B_{\mu}$ mixing term in
the low--energy effective Lagrangian.

 The full Lagrangian of SBTC, after introducing auxiliary fields, is:
\bea
{\cal{L}} & = & \bar{\psi} \left( \gamma^{\mu} D_{\mu}
           - \Phi \frac{1+\gamma^{5}}{2}
           -\Phi^{\dagger} \frac{1-\gamma^{5}}{2} \right)
          \psi
\nonumber \\
& & - \frac{M^{2}}{2 g^{2}} \tr (\Phi\Phi^{\dagger})
         - \frac{M^{2}}{g^{2}} \tr( V_{\mu} V^{\mu} + A_{\mu} A^{\mu} ),
\eea
 where:
\begin{eqnarray}
 D_{\mu} & = & i \partial_{\mu}
           - V_{\mu}-A_{\mu} \gamma^{5}
          - g_{1} B_{\mu}
              \frac{1+\gamma_{5}}{2}
           - g_{2} W_{\mu}
              \frac{1-\gamma_{5}}{2}.
\end{eqnarray}
 For the sake of brevity we write
\begin{eqnarray}
\Phi &\equiv & \sum_{\alpha=0}^{3} \Phi^{\alpha} \frac{\tau^{\alpha}}{2},
\qquad
V_{\mu} \equiv \sum_{\alpha=0}^{3} V^{\alpha}_{\mu} \frac{\tau^{\alpha}}{2},
\qquad
A_{\mu} \equiv \sum_{\alpha=0}^{3} A^{\alpha}_{\mu} \frac{\tau^{\alpha}}{2},
\nonumber \\
W_{\mu} &\equiv & \sum_{a=1}^{3} W^{a}_{\mu} \frac{\tau^{a}}{2},
\qquad
B_{\mu} \equiv B_{\mu} \frac{\tau^{3}}{2},
\end{eqnarray}
where $\tau^{0}={\bf 1}$.  $g_1$ and $g_2$ are respectively the $U(1)_Y$
and $SU(2)_L$ coupling constants.

It is convenient to make a shift of variables as follows:
\begin{eqnarray}
V_{\mu} & \longrightarrow & V_{\mu}
             - \frac{g_{1}}{2} B_{\mu}
             - \frac{g_{2}}{2} W_{\mu}  \nonumber \\
A_{\mu} & \longrightarrow & A_{\mu}
             - \frac{g_{1}}{2} B_{\mu}
            + \frac{g_{2}}{2} W_{\mu}.
\end{eqnarray}
After shifting the fields and integrating out the fermion degrees of freedom,
the effective action becomes:
\begin{eqnarray}
 {I} & = & - i N_{TC} \Tr\; \ln \left( i \gamma^{\mu} D_{\mu}
           - \Phi \frac{1+\gamma^{5}}{2}
           -\Phi^{\dagger} \frac{1-\gamma^{5}}{2} \right)    \nonumber  \\
   &   & + \int d^{4} x \;
       \left( - \frac{M^{2}}{2 g^{2}} \tr (\Phi\Phi^{\dagger})
         - \frac{M^{2}}{g^{2}}
     \tr  ( V_{\mu}
             - \frac{g_{1}}{2} B_{\mu}
             - \frac{g_{2}}{2} W_{\mu} )^{2} \right. \nonumber \\
   &    &  \qquad \left.   +  \tr ( A_{\mu}
             - \frac{g_{1}}{2} B_{\mu}
             + \frac{g_{2}}{2} W_{\mu}  )^{2} \right).
\end{eqnarray}
We obtain the kinetic term, cubic interaction term, etc.,  of the effective
action by expanding the above expression in terms of the fields around the
vacuum. The vacuum expectation value of $\Phi = m$ satisfies the gap equation:
\beq
       4 N_{TC}  \int \frac{d^{4}p}{(2\pi)^{4}} \frac{1}{p^{2}-m^{2}}
            = \frac{M^{2}}{g^{2}}.
\eeq
Using the auxiliary field method, the leading term in the $1/N_{TC}$ expansion
is obtained by carrying out the fermion one--loop calculation.
The kinetic term which is relevant to calculating
the $S$ parameter is:
\begin{eqnarray}
 I_{kin} & = & \int \frac{d^{4} q}{(2\pi)^{4}}    \nonumber \\
&&\left[
  \tr( W_{\mu}(q)W_{\nu}(-q) ) (-q^{2}g^{\mu\nu} + q^{\mu}q^{\nu} )\right.
\nonumber \\
&& + \tr( B_{\mu}(q)B_{\nu}(-q) ) (-q^{2}g^{\mu\nu} + q^{\mu}q^{\nu} )
\nonumber \\
&& + \tr( V_{\mu}(q)V_{\nu}(-q) ) (-q^{2}g^{\mu\nu} + q^{\mu}q^{\nu} )
   c_{1}(q^{2})
\nonumber \\
&& + \tr( A_{\mu}(q)A_{\nu}(-q) ) ((-q^{2}g^{\mu\nu} + q^{\mu}q^{\nu} )
   c_{1}(q^{2}) + g^{\mu\nu} c_{2}(q^{2})m^{2})
\nonumber \\
&& \left. + \frac{M^{2}}{g^{2}}
   (  \tr(V_{\mu}-\frac{g_{1}}{2}B_{\mu}-\frac{g_{2}}{2}W_{\mu})^{2}
    + \tr(A_{\mu}-\frac{g_{1}}{2}B_{\mu}+\frac{g_{2}}{2}W_{\mu})^{2}) \right].
\end{eqnarray}
Here:
\begin{eqnarray}
c_{1}(q^{2}) & = & \frac{N_{TC}}{(2\pi)^{2}}\int_{0}^{1} dx\;
               x(1-x)\ln[{M^{2}}/(m^{2}-x(1-x)q^{2})]
\nonumber \\
c_{2}(q^{2}) & = & \frac{N_{TC}}{(2\pi)^{2}}\int_{0}^{1} dx \;
               \ln[{M^{2}}/({m^{2}-x(1-x)q^{2}})].
\end{eqnarray}
We have used the unitary gauge so that the Nambu--Goldstone bosons
are already abosorbed into the axial--vector fields.
The kinetic terms for Higgs and charged scalars are omitted because
they are irrelevant to the $S$ parameter.

In order to obtain the low--energy effective action, one has to eliminate
techni-$\rho$ and techni-$A_{1}$ by using the equations of motion.
The solutions of the equations of motion for the previous action are,
\begin{eqnarray}
V^{a}_{\mu} & = &
\left( \frac{-q^{2}g_{\mu\nu}+q_{\mu}q_{\nu}}
      {q^{2}- m_{\rho}^{2}}
  +  g_{\mu\nu} \right) \frac{g_{2}}{2} W^{a \nu}  \nonumber \\
A^{a}_{\mu} & = &
- \frac{m_{\rho}^{2}}{m_{A_{1}}^{2}}
\left( \frac{-q^{2}g_{\mu\nu}+q_{\mu}q_{\nu}}
      {q^{2}-m_{A_{1}}^{2}}
  + g_{\mu\nu}
       \right) \frac{g_{2}}{2} W^{a \nu},
\end{eqnarray}
for $a=1,2$,  and:
\begin{eqnarray}
V^{3}_{\mu} & = &
\left( \frac{-q^{2}g_{\mu\nu}+q_{\mu}q_{\nu}}
      {q^{2}- m_{\rho}^{2}}
  +  g_{\mu\nu} \right)
(\frac{g_{1}}{2} B^{\nu}
 + \frac{g_{2}}{2} W^{3 \nu})
\nonumber \\
A^{3}_{\mu} & = &
- \frac{m_{\rho}^{2}}{m_{A_{1}}^{2}}
\left( \frac{-q^{2}g_{\mu\nu}+q_{\mu}q_{\nu}}
      {q^{2}-m_{A_{1}}^{2}}
  + g_{\mu\nu}
       \right)
(\frac{g_{1}}{2} B^{\nu}
 -\frac{g_{2}}{2} W^{3\nu}),
\end{eqnarray}
where:
\begin{eqnarray}
  m_{\rho}^{2} & = & \frac{N_{TC}M^{2}}{c_{1}g^{2}}  \nonumber \\
  m_{A_{1}}^{2} & = & \frac{c_{2}m^{2}}{c_{1}} + m_{\rho}^{2}.
\end{eqnarray}
Substituting this expression into eq.(44),
\begin{eqnarray}
I_{kin} & = &
 \frac{1}{2} \int \frac{d^{4}q}{(2\pi)^{4}}\nonumber \\
& & \left[ W^{a}_{\mu}W^{a}_{\nu} [
(-q^{2}g^{\mu\nu}+q^{\mu}q^{\nu})
-\left(\frac{g^2_{2}c_1}{4}\right)
\frac{m_{\rho}^{2}}{q^{2}-m_{\rho}^{2}}
(-q^{2}g^{\mu\nu}+q^{\mu}q^{\nu}) \right.
\nonumber \\
& & +
\left(\frac{g^2_{2}}{4}\right)\frac{m_{\rho}^{2}}{m_{A_{1}}^{2}}
(c_{2}m^{2} g^{\mu\nu}
+ c_{1}\frac{m_{\rho}^{2}}{q^{2}-m_{A_{1}}^{2}}
(-q^{2}g^{\mu\nu}+q^{\mu}q^{\nu})) ]
\nonumber \\
 & & +  B_{\mu}B_{\nu}
[(-q^{2}g^{\mu\nu}+q^{\mu}q^{\nu})
-\left(\frac{g^2_{1}c_1}{4}\right)\frac{m_{\rho}^{2}}{q^{2}-  m_{\rho}^{2}}
(-q^{2}g^{\mu\nu}+q^{\mu}q^{\nu})
\nonumber \\
 & & + \left(\frac{g^2_{1}}{4}\right)\frac{m_{\rho}^{2}}{m_{A_{1}}^{2}}
(c_{2}m^{2} g^{\mu\nu}
+ c_{1}\frac{m_{\rho}^{2}}{q^{2}-m_{A_{1}}^{2}}
(-q^{2}g^{\mu\nu}+q^{\mu}q^{\nu})) ]
\nonumber \\
& & - \frac{g_{1}g_2 }{2}  B_{\mu}W^{3}_{\nu}
[\frac{m_{\rho}^{2}}{q^{2}-m_{\rho}^{2}} c_1(-q^{2}g^{\mu\nu}+q^{\mu}q^{\nu})
\
\nonumber \\
& & \left. + \frac{m_{\rho}^{2}}{m_{A_{1}}^{2}}
    (c_{2}m^{2} g^{\mu\nu}
-c_1\frac{m_{\rho}^{2}}{q^{2}-m_{A_{1}}^{2}}
(-q^{2}g^{\mu\nu}+q^{\mu}q^{\nu})
     )]\right].
\end{eqnarray}

Now we can compute the  $S$ parameter.
{}From eq.(19) the $S$ parameter can be
written as:
\beq
     S = -16\pi \Pi_{3Y}^{\prime }(0) ,
\eeq
where  $\Pi_{3Y}$ is
just the coefficient of the $g_{1}g_{2}B_{\mu}W^{3 \mu}$ term
in the effective action. We therefore have:
\beq
  \Pi_{3Y} = \frac{1}{4}
   \left[\frac{m_{\rho}^{2}}{q^{2}-m_{\rho}^{2}}c_{1}q^{2}
           -\frac{m_{\rho}^{2}}{m_{A_{1}}^{2}}
                (c_{2}m^{2}+
              \frac{m_{\rho}^{2}}{q^{2}-m_{A_{1}}^{2}}c_{1}q^{2})
              \right].
\eeq
It can also be shown that the first term is $\Pi_{VV}$
and the second term is $-\Pi_{AA}$.

Substituting this expression and eq.(45) into  eq.(50),
we obtain
\beq
              S =  \frac{N_{TC}}{6\pi}
                \left( 1 + ( \ln\frac{M^{2}}{m^{2}} - 1 )
                  ( 1- \frac{1}{\eta^{2}} ) \right)
\eeq
which is as in eq.(37) with $\eta = m_{A_1}^2/m_\rho^2$.

\vskip .1in
\noindent
{\bf IV. Breaking of Technicolor}
\vskip .1in

How is the technicolor interaction broken?  The breaking can either involve
additional interactions, a somewhat epicyclic scenario, or it may
occur in certain gauge theories automatically.
While we have no compelling self-destructing theories in mind, they can
certainly exist (``tumbling'' [15]).
The unitary, orthogonal and exceptional Lie groups have
complex (chiral) representations and thus in principle can break themselves.
(Symplectic groups are real.)  Unfortunately, there are no chiral gauge
theories, at least with minimal fermion content, that break themselves
completely or become null in the sense defined earlier [16].

As a minimal toy example, consider a chiral $SU(3)$ model.  Chiral
asymptotically free, anomaly--free $SU(N)$ theories were classifed by Eichten,
Kang and Koh [17].  The $SU(3)$ model requires seven ${\bf\bar 3}$ and one
${\bf
6}$ representations.  The degenerate most attractive channels for condensation
are the ${\bf 3}$ of the
${\bf\bar 3}\otimes{\bf 6}$ and the ${\bf 6}$ of the ${\bf 6}\otimes{\bf 6}.$
There is, after instanton effects are taken into account, a global symmetry
of $SU(7)\times U(1).$  The electroweak theory is embedded by assembling four
${\bf\bar 3}$ into two electroweak doublets with hypercharges $Y$ = $\pm$1,
leaving the remaining three ${\bf\bar 3}$ and the ${\bf 6}$ as electroweak
singlets.  The hypercharge group can be embedded in an $SU(2)_R$ symmetry.
The theory has no electroweak, TC or mixed anomalies.
The electroweak corrections to the effective potential of the vacuum favor the
${\bf 3}$ of the ${\bf\bar 3}\otimes{\bf 6}$ as the most attractive channel,
where the ${\bf\bar 3}$ are the electroweak
doublets, thus preserving the $SU(2)$ custodial symmetry.
The $SU(3)_{TC}$ is broken to $SU(2)_{TC}$ by this vacuum, and five of the
eight
technigluons acquire mass.  The electroweak
$SU(2)_L$ and the $U(1)_Y$ are broken, as neither is a vector symmetry.
The $U(1)_Y$ is not vector-like, in spite of the hypercharge assignments,
because the electroweak doublets are not vector-like under the $SU(3)_{TC}.$
The electromagnetic $U(1)_Q$ subgroup is unbroken,
because it {\it is} vector-like --- the remaining $SU(2)_{TC}$
is pseudoreal, and the ${\bf 2}$ and ${\bf\bar 2}$ representations can be
assembled to form Dirac fermions.  The Vafa--Witten theorem again applies here,
forbidding the dynamical breaking of vector symmetries.

Applying the NJL techniques and results of section~II to this model results in
a mass gap of $m\simeq$ 720 GeV and a technigluon mass of $M\simeq$ 770 GeV,
with $\eta$ = 3.  Since the technicondensates break both the technicolor
and electroweak groups, the TC coupling is $g(M)$ = 2$\cdot$770 GeV/246 GeV
$\simeq$ 6.3.  This small hierarchy is at the edge of the NJL model's
reliability (and the argument of the log has been taken to
be $1+M/m$ to allow larger $\eta$).

After the electroweak embedding, the global symmetry is broken to
$SU(3)\times SU(2)_L\times [U(1)]^3;$ after spontaneous symmetry breakdown,
a global $SU(3)\times U(1)$ remains and
there are three Goldstone bosons
eaten by the $W$ and $Z,$ two EW singlet true Goldstones, and eight
pseudoGoldstones of mass 150--200 GeV.
The Higgses behave,
below the cutoff, like a two-doublet Higgs theory [18].  The charged and
light neutral Higgses have mass of about 100 GeV, the pseudoscalar ``axion''
a somewhat lighter mass, and the heavy neutral Higgs a mass of about 900
GeV.  There is also a scalar isotriplet.  The technifermions acquire
dynamical Dirac masses
coupling the ${\bf\bar 3}$ electroweak doublets
and the ${\bf 6}.$  There are one massless
electroweak/$SU(2)_{TC}$ singlet and two electroweak/$SU(2)_{TC}$ doublets
of mass $\sqrt{2}$(720 GeV) = 1020 GeV.  The $SU(2)_{TC}$ confines the
latter at 200 GeV.
We can estimate the $S$ from fermion loops
using the bubble formula eq.(37).  For two doublets, $S = (2/3\pi )(0.79)$ =
0.17.  From $SU(2)_{TC}$
technigluon exchange across the loop, there are perturbative
corrections $\sim$ 30\%, but impossible to compute at only
lowest order [19].  The non-perturbative $SU(2)_{TC}$
corrections can be estimated using the methods of
Pagels and Stokar [20] to be negligible.
There are scalar (Higgs and pseudoGoldstone) contributions to $S$ which
are next order in
$1/N_{TC}$, as well as higher order corrections in $g_2^2$, and we
have no systematic way to include all other corrections of this
order.
The estimate
$S\approx 0.17$ is probably too large for current electroweak limits.  There
are
additional condensates.  The $SU(2)_{TC}$ is pseudoreal and thus cannot
break itself, but the left-over global $SU(3)\times U(1)$ is broken completely,
with nine additional electroweak singlet true Goldstones.  This $SU(3)$
model, while not realistic, may be taken as a kind of optimal case, inasmuch
as the model is self-contained, and the remnant
$SU(2)_{TC},$ while not null, is the smallest non-Abelian group.
The problem with self--contained models of this kind is clear: they
require too many degrees of freedom contributing to $S$.

Ultimately, it is necessary to maintain the smallest number of additional
degrees of freedom possible
in order to minimize $S$. Thus, a fourth generation  or a
top--condensate  scheme  may have some advantages. The arguments presented here
do not apply
directly to a ``topcolor'' model [10] since there one has a maximally broken
custodial $SU(2)$.  There we expect the formula for $S$ to resemble eq.(29),
in which case negative $S$ can readily occur.  However, the $T$ constraint
would seem to disfavor such a model since evidently $m_{top}$ cannot be
as large as one would expect if the new physics is at $\sim 1 $ TeV.
Nevertheless, it is interesting to ask how large the effects of the virtual
resonances are on the $T$ parameter, since one might expect a
suppression of $T$ for large $m_{top}$ in such a scheme.  An investigation
of this question is underway.

\vskip .1in
\noindent
{\bf V. Conclusions}
\vskip .1in

In Figure 1 we plot the result for $S$ as obtained in the
fermion bubble approximation to our model of eq.(4).
Here we use eq.(37) together with the gap equation constraint, eq.(35),
upon  the parameter $\eta$.  The gap equation implies that
$\eta $ can be viewed as a function of the mass ratio $M/m$, so it is
also plotted.
The one-doublet
free-fermion loop contribution corresponds to $6\pi S/N_{TC} = 1$.
We see that the effect of the virtual vector technimesons is to
increase $S$ slightly for strong coupling.  The approximations of the
model are not valid for $\ln(M^2/m^2) \lta 2$, for we then become
sensitive to higher orders in $m^2/M^2$. For example, the
maximum value of $\eta$ permitted, such that the gap equation
has a nontrivial solution, depends upon the argument of the logarithm.  For
a logarithm of the form $\ln (M^2/m^2)$ we find $\eta \lta 1.6$,
while the form $\ln (M^2/m^2 + 1)$ we find $\eta \lta 3.3$.  The
physical difference between these forms are terms of higher order
in $m^2/M^2$, and we cannot reliably compute their effects.  Indeed,
in writing eq.(4) as the effective Lagrangian, we have already discarded
the higher order terms in an operator product expansion of the
effects of single massive technigluon exchange.

Our model approximates all of the dynamics as pure $s$--wave.  Obviously,
a confining theory has strong higher partial waves, and we would expect
these to contribute to $S$.  The results of Peskin and Takeuchi [6], which
reflect the full confinement effects of QCD through
the saturation of the $\Pi_{VV}-\Pi_{AA}$ by
the real $\rho$ and $A_1$ resonances, yield $6\pi S/N_{TC}\approx 2$.
The pure $s$--wave result of our analysis of virtual $\rho$ and $A_1$
resonances is $6\pi S/N_{TC}\approx 1.5$ when $\eta$
is maximal.  Thus our results do capture some
of the effects of a full QCD--like theory.  The key point of our
present analysis is that we may choose $\eta\rightarrow 1$ without
drastic fine--tuning, and suppress $S$ somewhat
toward the free-fermion loop
result.  For example, choosing $M/m \sim 10$ yields $6\pi S/N_{TC}
\sim 1.3$ and $\eta \sim 1.05$.  The behavior of the function
of eq.(37) also allows a reduction of $S$ for $\eta \gta 1.4$
with $M/m \lta 3$, though this is a less reliable
limit of the approximation.
(Note that in the extreme example of
the top condensate scheme where $M/m$ is fine--tuned to $\sim 10^{15}$
we see that there is negligible virtual resonance effects upon
$S$.)

Spontaneously broken technicolor thus offers a mild advantage in reducing the
unwanted resonance contributions to $S$.
In estimating $S$ in an SBTC theory one can rely upon the free--fermion
loop result of $N_{TC}/6\pi$ if one is willing to tolerate
some fine--tuning near to the critical coupling regime.  We
emphasize thatwe  have not given
a method for engineering a model with negative $S$.

There is no good reason,
in light of the strong ETC gymnastics that are required to maintain
small flavor--changing neutral current processes and a heavy top quark
[3, 4], to argue that technicolor need be an unbroken,
QCD--like theory with confined techniquarks.  A strong, broken
gauge theory would be a novelty in nature, but there is no reason to
rule out the possibility.  For electroweak symmetry breaking
it is clearly necessary to maintain the smallest number of
degrees of freedom possible
in order to minimize $S$. A fourth generation scheme, such as
the model of ref.[11],  or a
top--condensate  scheme [9, 10], thus has clear advantages, and we
have briefly discussed others.  This is obviously not an exhaustive list,
and there may prove to be other advantages to model building in which
technicolor, together with its extension, is spontaneously broken.

\vskip .1in
\noindent
{\bf Acknowledgments}
\vskip .1in

The authors would like to thank Bill Bardeen, Estia Eichten, Dirk
Jungnickel, and Tatsu Takeuchi of Fermilab
for valuable discussions.

\vskip 1.0in
\noindent
{\bf Figure Caption}
\vskip .1in
\noindent
1. The $S$ parameter, as obtained in the
fermion bubble approximation (solid line), eq.(37)
is plotted against $\ln(M^2/m^2)$.
Here the gap equation constraint, eq.(35), is implemented
for the parameter $\eta$, which is also plotted
(dashed line).

\noindent
{\bf References}
\begin{enumerate}
\item
S.~Weinberg, {\it Phys. Rev.} {\bf D19}, 1277 (1979);\\
L.~Susskind, {\it Phys. Rev.} {\bf D20}, 2619 (1979).
\item
S.~Dimopoulos, L.~Susskind, {\it Nucl. Phys.} {\bf B155}, 237 (1979);\\
E.~Eichten, K.~Lane, {\it Phys. Lett.} {\bf 90B}, 125 (1980).
\item
B.~Holdom, {\it Phys. Lett.} {\bf B150}, 301 (1985); {\bf B198}, 535 (1987);
T.~Appelquist, D.~Karabli, L.~C.~R.~Wijewardhana, {\it Phys. Rev. Lett.}
{\bf 57}, 957 (1986); M. Bando, \etal, {\em Phys. Rev. Lett.}
{\bf 59} 389 (1987) ; V. Miransky, {\em Nuovo Cimento} {\bf 90A}
 1 (1985).
\item
T. Appelquist, T. Takeuchi, M. Einhorn, L.C.R. Wijewardhana,
{\it Phys. Lett.} {\bf B220},  223 (1989); R. Mendel, V. Miransky,
{\it Phys. Lett.} {\bf B286}, 384 (1991).
\item
D.~C.~Kennedy, B.~W.~Lynn, {\it Nucl. Phys.} {\bf B322}, 1 (1989);\\
D.~C.~Kennedy, {\it Phys. Lett.} {\bf B268}, 86 (1991).
\item
M.~E.~Peskin, T.~Takeuchi, {\it Phys. Rev.} {\bf D46}, 381 (1992);\\
T.~Takeuchi, private communication.
\item
P.~G.~Langacker, U.~Pennsylvania preprint UPR-0492T (1992),\\
and private communication.
\item
R.~Sundrum, S.~D.~H.~Hsu, Berkeley preprint UCB-PTH-91-34 (1991, revised 1992).
\item
W.~A. Bardeen, C.~T. Hill, M.~Lindner, {\it Phys. Rev.} {\bf D41}, 1647 (1990),
 \\ and references therein.
 \item
C.~T.~Hill, {\it Phys. Lett.} {\bf B266}, 419 (1991);\\
S. Martin, {\it Phys. Rev.} {\bf D46}, 2197 (1992) and
{\it Phys. Rev.} {\bf D45}, 4283 (1992);\\
N. Evans, S. King, D. Ross, ``Top Quark Condensation from
Broken Family Symmetry,''  Southampton Univ. preprint,
SHEP--91--92--11 (1992).
\item
C.~T.~Hill, M.~Luty, E.~A.~Paschos, {\it Phys. Rev.} {\bf D43}, 3011 (1991);
T. Elliot, S. F. King,  {\it Phys. Lett.} {\bf B283}, 371 (1992);
\item
Y.~Nambu, G.~Jona-Lasinio, {\it Phys. Rev.} {\bf 122}, 345 (1961);
 {\bf 124}, 246 (1961);\\
 B.~Rosenstein, B.~Warr, S.~H.~Park, {\it Phys. Rep.} {\bf 205}, 59 (1991);\\
 S.~P.~Klevansky, {\it Rev. Mod. Phys.} {\bf 64}, 649 (1992).
\item
C.~Vafa, E.~Witten, {\it Nucl. Phys.} {\bf B234}, 173 (1984).
\item
J.~Goldstone, {\it Nuovo Cim.} {\bf 19}, 154 (1961).
\item
S.~Raby, S.~Dimopoulos, L.~Susskind, {\it Nucl. Phys.} {\bf B169}, 373 (1980).
\item
R.~Slansky, {\it Phys. Rep.} {\bf 79}, 1 (1981).
\item
E.~Eichten, K.~Kang, I.-G.~Koh, {\it J.~Math.~Phys.} {\bf 23}, 2529 (1982);\\
 E.~Eichten, R.~Peccei, J.~Preskill, D.~Zeppenfeld, {\it Nucl. Phys.} {\bf
 B268}, 161 (1986).
\item
J.~F.~Gunion, H.~E.~Haber, G.~Kane, S.~Dawson, {\it The Higgs Hunter's Guide}
 (Redwood City, California: Addison-Wesley, 1990) chapter 4;\\
 H.~E.~Haber, A.~Pomarol, UCSC SCIPP preprint 92/29 (1992);\\
 H.~E.~Haber, UCSC SCIPP preprint 92/31 (1992);\\
 C.~T.~Hill, C.~N.~Leung, S.~Rao, {\it Nucl. Phys.} {\bf B262}, 517 (1985).
\item
A.~Djouadi, C.~Verzegnassi, {\it Phys. Lett.} {\bf B195}, 265 (1987);\\
 A.~Djouadi, {\it Nuovo Cim.} {\bf 100A}, 357 (1988);\\
 B.~A.~Kniehl, {\it Nucl. Phys.} {\bf B347}, 86 (1990);\\
 F.~Halzen, B.~A.~Kniehl, {\it Nucl. Phys.} {\bf B353}, 567 (1991).
\item
H.~Pagels, {\it Phys. Rev} {\bf D19}, 3080 (1979);\\
 H.~Pagels, S.~Stokar, {\it Phys. Rev.} {\bf D20}, 2947 (1979).
\end{enumerate}

\end{document}